\documentclass[english,apl,twocolumn,amsmath,amssymb]{revtex4}
\usepackage[T1]{fontenc}
\usepackage[latin9]{inputenc}
\usepackage{amsmath}
\usepackage{graphicx}
\usepackage{esint}
\usepackage{babel}

\begin{document}

\title{Nonlinear Electric Metamaterials}

\author{David A. Powell}
\email{david.a.powell@anu.edu.au}

\author{Ilya V. Shadrivov}
\author{Yuri S. Kivshar}

\affiliation{Nonlinear Physics Center, Research School of Physics and Engineering,\\
 Australian National University, Canberra ACT 0200, Australia }

\begin{abstract}
We propose and design a new type of nonlinear metamaterials exhibiting a
resonant electric response at microwave frequencies. By introducing
a varactor diode as a nonlinear element within each resonator, we
are able to shift the frequency of the electric mode stop-band by
changing the incident power, without affecting the magnetic response.
These elements could be combined with the previously developed nonlinear 
magnetic metamaterials in order to create negative
index media with control over both electric and magnetic nonlinearities.
\end{abstract}

\maketitle

A negative index metamaterials usually consists of a composite structure
which simultaneously exhibits negative electric and magnetic responses
over some frequency band~\cite{Veselago1968}. Split-ring resonators
(SRRs) are the best known elements of structures with a negative magnetic
response, and arrays of long wires are commonly used to create a negative
electric permittivity~\cite{Smith2000}. This approach has the advantage
that the broad negative electric response of the wires can relatively
easily be overlapped with the narrow-band negative magnetic response
of the SRRs. However, long wires are unsuitable for many applications,
such as transform optics which require local variations in the unit
cell parameters and which may require a non-periodic arrangement of
unit cells. One of the solutions to this problem is the use of electric
resonators constructed from two resonant loops having a fundamental
mode whereby the total magnetic dipole moment is canceled out, leaving
a nonvanishing electric dipole moment~\cite{Liu2007,Schurig2006}.
Such structures have been demonstrated in the microwave and THz bands,
and have been shown to have their response tuned by an optical pump
signal~\cite{Chen2007c}.

The ability to engineer the linear response of metamaterials to achieve
exotic values of permittivity and permeability is well known, however
their strong field localization also makes them very well suited to
exhibiting exotic nonlinear phenomena~\cite{Gorkunov2004,Pendry1999,Zharov2003}.
This is due to the strong increase in field strength at the resonant
frequency and also the local field strength which can be much larger
than the average incident field strength. Previous work has shown
that individual split ring resonators can be made tunable and nonlinear
by the introduction of diodes with voltage-controlled capacitance
\cite{Gil2004,Shadrivov2006a,Powell2007}, and that such resonators
can be combined to create bulk nonlinear metamaterial structures~\cite{Shadrivov2008a,Shadrivov2008}.

\begin{figure}
\includegraphics[width=1\columnwidth]{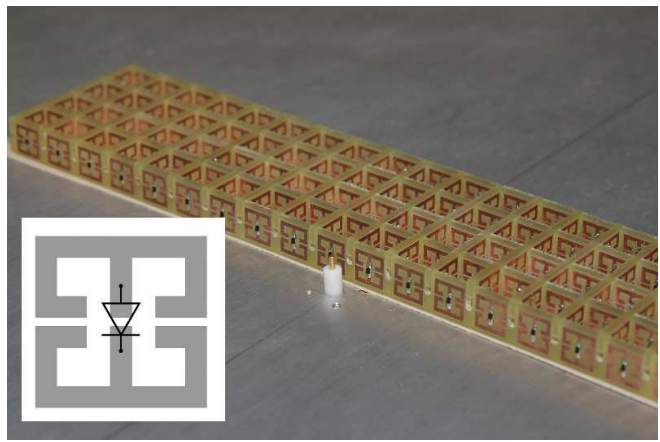}
\caption{Slab of nonlinear electric metamaterials used in our experiments.
The inset shows one structural element.
\label{fig:structure}}
\end{figure}

Given that the nonlinear shift in resonance results in a very strong
nonlinear magnetic response for SRRs, we take a similar approach to
the design of nonlinear electric resonators, in order to obtain a
strong nonlinear electric response. Our structure is shown in Fig.~\ref{fig:structure},
where two perpendicular sets of boards are introduced to create a
relatively isotropic response. Within each resonator, an additional
gap is introduced where a varactor diode is placed, introducing additional
series capacitance in order to tune the resonant frequency. The lattice
period is 11mm, and the resonators are fabricated on copper-clad FR4
with width and height of 8mm, track width of 1mm, with outer gaps
of 0.4mm separation and 2.4mm length.

\begin{figure}
\includegraphics[width=1\columnwidth]{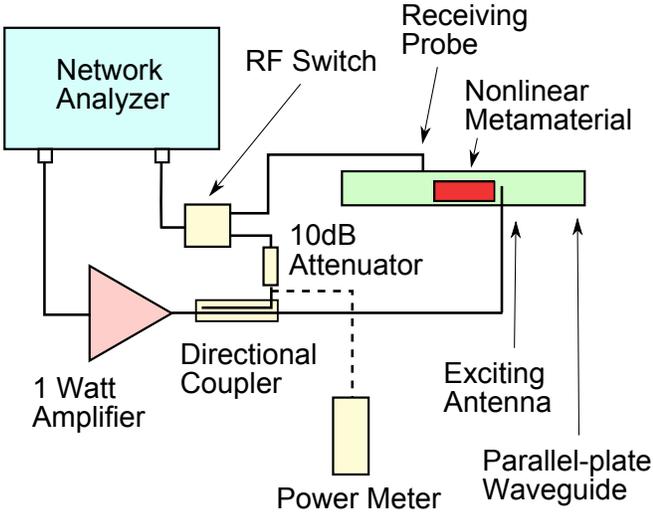}
\caption{Measurement configuration~\label{fig:measurement-setup}}
\end{figure}

The sample is placed in a parallel plate waveguide and is excited
by a vector network analyzer via a monopole probe, with a microwave
amplifier included in the system in order to increase the available
excitation power, as shown in Fig.~\ref{fig:measurement-setup}.
As there is significant ripple and some gain compression in the transmission
response of the amplifier, a power calibration is first performed
for each desired power level by connecting a power meter to the directional
coupler, as shown by the dotted line. In order to know the value of
the fields fed to the waveguide, the output of the amplifier is sampled
using a directional coupler. The total transmission through the amplifier,
waveguide, and metamaterial is then normalized to this quantity, which
has largely eliminated artifacts due to the amplifier response.

\begin{figure}
\begin{centering}
\includegraphics[width=1\columnwidth]{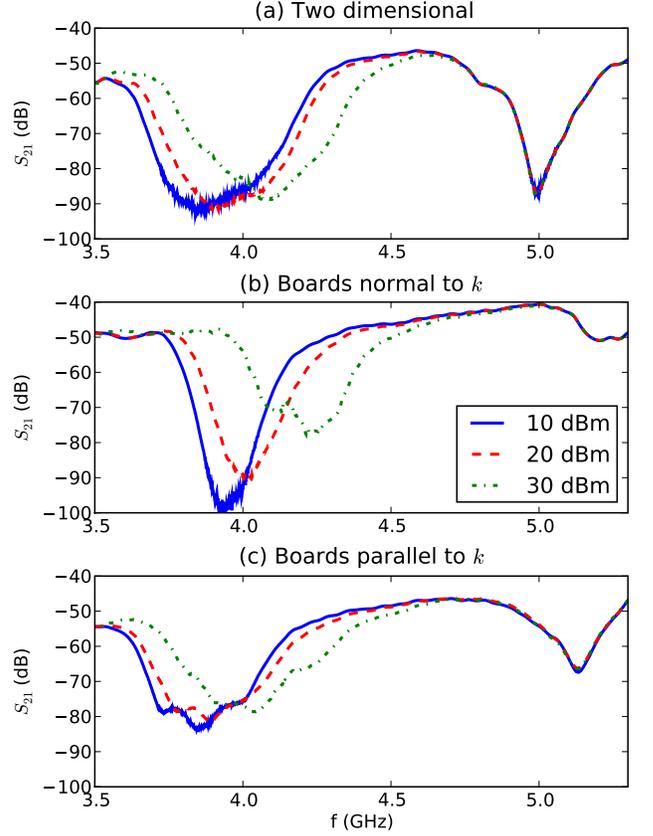}
\par\end{centering}
\caption{(a-c) Experimentally measured nonlinear transmission response at incident powers of 10, 20
and 30dBm for different orientations.
\label{fig:transmission}}
\end{figure}

We measure the transmission response at incident powers of 10, 20
and 30dBm as shown in Fig.~\ref{fig:transmission}(a). At the lowest
incident power (10dBm), there is negligible tuning of the response
by the incident wave, thus the transmission response in this case
is essentially linear. The large insertion loss away from resonance
is due to the deliberately mismatched receiving probe which is designed
for minimal perturbation of the fields within the structure.

\begin{figure}
\begin{centering}
\includegraphics[bb=0bp 20mm 640bp 320bp,width=1\columnwidth]{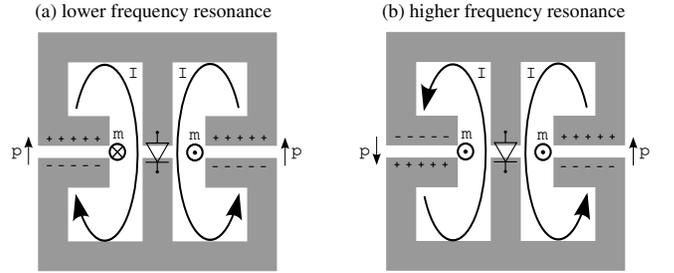}
\caption{Resonant modes of the electric resonators, showing the two current loops 
$I$ and their electric and magnetic dipole moments $p$, and $m$.
\label{fig:modes}}
\par\end{centering}
\end{figure}

It can be seen that our structure exhibits two resonant stop bands
within the measured frequency range. The lower band is due to currents
flowing in opposite directions in each of the two rings of the resonator,
as shown in Fig.~\ref{fig:modes}(a). This results in two equal magnetic
dipoles of opposite orientation, which have a zero net magnetic dipole
moment%
\footnote{For most orientations of the resonators relative to the incident wave-vector,
retardation effects will cause some phase-shift in the fields across
each element, thus complete cancelation of the dipole moment may
not occur at the resonant frequency.%
}, and a dominant electric dipole moment due to the identical charge
accumulation across each gap. Importantly, there is significant net
current flowing through the central conductor containing the varactor
diode. Thus we see that this resonant frequency is strongly modified
by the incident power.

On the other hand, Fig.~\ref{fig:transmission}(a) shows that the
higher frequency mode does not shift its frequency with a change of
the incident power. As can be seen in Fig.~\ref{fig:modes}(b), this
mode consists of two current loops flowing in the same direction,
thus their magnetic dipole moments add constructively. As the accumulated
charges across the gaps have opposite directions, this results in
a vanishing electric dipole moment. As there is no net current through
the central conductor, the nonlinear response of the varactor diode
does not come into play. These mode configurations, and the stop-band
locations, are confirmed by numerical simulations of transmission
performed in CST Microwave Studio \cite{CST} using a single element
with electric boundaries in the vertical direction and magnetic boundaries
in the horizontal direction.

For comparison, we investigate the two different orientations of the
circuit boards individually, i.e. those which are normal to the direction
of propagation, and those which are parallel to it. The nonlinear
transmission responses for these structures are shown in Figs.~\ref{fig:transmission}(b)
and (c) respectively. In both cases a significant nonlinear response
still occurs. In the case of the boards being perpendicular to the
direction of propagation, the higher frequency magnetic stop-band
does not exist. This is due to the symmetry of the fields across the
gaps and the lack of any magnetic field component normal to the rings.
We note that as the experiment uses a cylindrically symmetric source
there is some component of the wavenumber normal to the nominal propagation
direction, hence some vestige of the second resonance remains. Also
of note is the fact that both resonances are noticeably modified in
the isotropic configuration compared to when they are measured separately.
This is likely to be due to the strong electrical interaction between
the nearest-neighbor boards in the orthogonal directions, due to their
gaps being in close proximity.

To verify the nature of metamaterial resonances, the most common approach
is to consider the reflection and transmission from a finite-thickness
slab, which are then inverted to find the equivalent refractive index
and impedance which will reconstruct the observed scattering. However
for most metamaterials reported in the literature, the permittivity
and permeability obtained by this method have non physical features,
such as violation of conservation of energy \cite{Simovski2007a}.
Thus we choose instead to calculate the electric and magnetic dipole
moments of the unit cell as a function of frequency, in order to verify
the intuitive picture of the nature of these resonances. We take the
following definitions of the electric and magnetic dipole moments
\cite{Raab2005} calculated over the unit cell of volume $V$ due
to current density $\bar{J}$ and charge density $\rho=-\nabla\cdot\bar{J}/j\omega$:

\begin{equation}
\bar{p}=\frac{1}{j\omega}\int_{V}\bar{J}\left(\bar{x}\right)d^{3}x\label{eq:electric_dipole}\end{equation}

\begin{equation}
\bar{m}=\frac{1}{2}\int_{V}\bar{x}\times\bar{J}\left(\bar{x}\right)d^{3}x\end{equation}

\begin{figure}
\includegraphics[width=1\columnwidth]{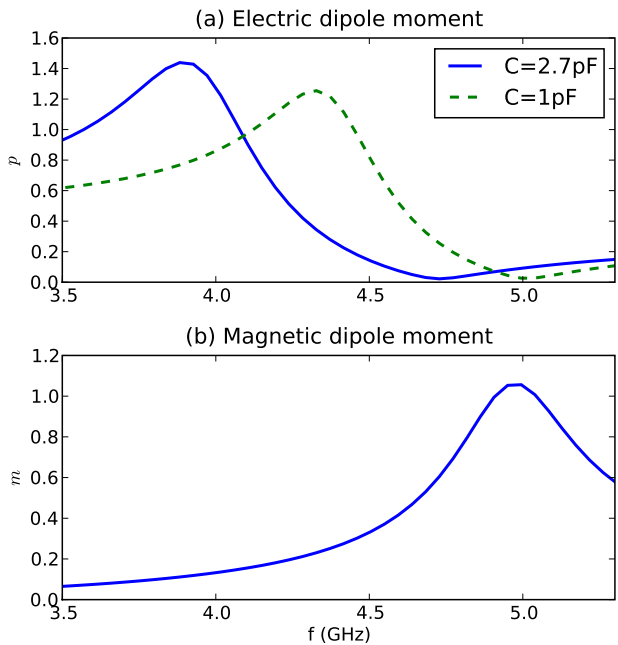}
\caption{Numerically calculated electric (a) and magnetic (b) dipole moments of the resonator aligned
parallel to $\bar{k}$ \label{fig:polsarisation}}
\end{figure}

From numerical simulations of a single layer of electric resonators,
we can readily retrieve the surface currents on the conductors, which
we assume to dominate over displacement currents within the dielectrics.
We also note that the currents flowing in the waveguide walls should
not be included in the calculation, as they represent the response
of the images of the unit cell, and thus should be considered as belonging
to virtual neighbors. Therefore in deriving Eq.~\eqref{eq:electric_dipole}
from the definition of the electric dipole moment based on the charge
distribution, we neglect the terms corresponding to currents flowing
on the boundaries of the unit cell.

With reference to Fig.~\ref{fig:modes}, the dominant electric dipole
moment will be in the vertical direction, and the dominant magnetic
dipole moment will be oriented out of the page and these components
are plotted in the normalized forms $p_{y}/E_{inc}\epsilon_{0}V$
and $m_{x}/H_{inc}V$ in Fig.~\ref{fig:polsarisation}. Here the
resonator is aligned parallel to the wave-vector as this is the simplest
case exhibiting both resonances. We have presented results for a capacitance
corresponding to the nominal unbiased diode capacitance of 2.7pF,
as well as a for the capacitance tuned to 1pF. This capacitance shift
is higher than what we expect to occur in the experiment due to nonlinear
self-tuning, however it clearly illustrates that a change in the loading
capacitance changes the resonant response of the electric dipole moment.
There is no visible difference between the two curves of the magnetic
dipole moment \emph{across the entire simulated frequency range},
again confirming that even in the case of very strong nonlinear or
external tuning the magnetic response remains constant and linear.

In conclusion, we have suggested, designed, and analyzed a new type of 
nonlinear metamaterial with a dominant negative electric response. We have 
showed that we are able to introduce nonlinearity
into the electric response making it tunable whilst leaving the magnetic 
response unchanged. We expect that our results would constitute the building blocks of
a complete nonlinear negative-index metamaterial containing both nonlinear
or tunable electric and magnetic elements which can be engineered
independently.


\end{document}